\documentstyle[twocolumn,aps]{revtex}

\begin{document}
\draft
\title{Entangled quantum clocks for measuring proper-time
difference
}
\author{WonYoung Hwang $^1$
\cite{email},
Doyeol (David) Ahn $^1$ \cite{byline},
Sung Woo Hwang $^1$ \cite{byline2}, and Yeong Deok Han $^2$ }

\address{$^1$ Institute of Quantum Information Processing
and Systems,
 University of Seoul, 90 Jeonnong, Tongdaemoon,
 Seoul 130-743, Korea
}
\address{$^2$ Division of Semiconductor, Electricity, and
Automobile Engineering,
Woosuk University, 490 Hujeong, Samrye, Wanju,
Cheonbuk 565-701, Korea
}
\maketitle
\begin{abstract}
We report that entangled pairs of
quantum clocks (non-degenerate quantum bits)
can be used as a specialized detector for precisely
measuring difference of proper-times that each constituent
quantum clock experiences. We describe
why the proposed scheme would be more precise
in the measurement of proper-time difference
than a scheme of two-separate-quantum-clocks.
We consider possibilities that the proposed
scheme can be used in precision test of
the relativity theory.
\end{abstract}
\pacs{03.67.-a, 06.30.Ft, 04.80.Cc }

\narrowtext

It is quantum entanglement that led to the historical
controversy over Einstein-Podolsky-Rosen experiment
\cite{eins} and then led to the Bell's inequality \cite{bell} that
explicitly revealed non-local nature of quantum mechanics.
On the other hand, entanglement is the key ingredient in
quantum information processing: for example, the speedup in
quantum computation \cite{shor} is obtained through the
parallel quantum operations on massively superposed states
which are entangled in general.
Recently, several new protocols using quantum entanglement
that have advantages over its classical counterparts
were proposed-
entanglement enhanced frequency measurement  \cite{boll},
quantum lithography \cite{boto,dang},
quantum clock synchronization based on shared prior entanglement
\cite{jozs,burt,shah,yurt,pres},
efficient quantum clock-transport scheme
\cite{chua}, and quantum enhanced positioning \cite{giov}.

In this paper, we propose a new application of
the entangled pairs of quantum clocks
(non-degenerate quantum bits)-
the specialized detector that precisely measures
difference of proper-times that each quantum clock experiences.
The proposed scheme is expected to be more precise in measuring
the proper-time difference than
a scheme where two separate quantum clocks are employed.
In this scheme, quantum clocks need to be accelerated for some
time-intervals and the acceleration's effects on quantum clocks
might be non-negligible. Thus appropriate handling of
the effects are necessary.
We suggest a solution and a utilization of this case.
Then we consider using the proposed scheme in the
precision test of relativistic time-dilation effects.

Let us assume that we have entangled pair of quantum clocks
in the state
\begin{equation}
\label{a}
|\Psi^- \rangle=
|0\rangle_A |1\rangle_B -|1\rangle_A |0\rangle_B,
\end{equation}
where $A$ and $B$ respectively
corresponds to each quantum clock whose proper-time
difference will be compared. 
(The normalization factor is omitted 
throughout this paper.)
We also assume that Hamiltonian ${\bf H}_\alpha$
for two mutually orthogonal states of a quantum clock,
$|0\rangle_\alpha$ and $|1\rangle_\alpha$
($\alpha=A, B$) is given by
\begin{equation}
 \label{b}
 {\bf H}_\alpha= E_\alpha \sigma_z,
\end{equation}
where ${\bf \sigma}_i$  ($i=x,y,z$)  is the Pauli operators.
The time evolution of each quantum clock is in general
given by the unitary operation
\begin{equation}
\label{c}
U_\alpha(t) |0\rangle_\alpha = e^{i E_\alpha t}  |0\rangle_\alpha,
\hspace{5mm}
U_\alpha(t) |1\rangle_\alpha = e^{-i E_\alpha t} |1\rangle_\alpha,
\end{equation}
where
$\hbar$ is set to be one.
When two clocks follow different space-time trajectories,
the time for each clock is given by it's own proper-time.
First let us consider the case $E_A=E_B=E$. (We will
later consider a general case where $E_\alpha$'s are
time-dependent and thus are not the same.)
After proper-times $t_A$ and $t_B$ have
elapsed for $A$ and $B$ quantum clocks, respectively,
the initial state of the quantum clocks in Eq. (\ref{a}) becomes
\begin{eqnarray}
\label{d}
&&U_A(t_A) U_B(t_B) |\Psi^- \rangle \nonumber\\
&=&
U_A(t_A)
(e^{-i E t_B}
|0\rangle_A |1\rangle_B
-e^{i E t_B}|1\rangle_A |0\rangle_B), \nonumber\\
&=&
 e^{-i E t_B} e^{i E t_A}
 |0\rangle_A |1\rangle_B
-e^{i E t_B} e^{-i E t_A}
 |1\rangle_A |0\rangle_B, \nonumber\\
&=&
 e^{iE\Delta t} |0\rangle_A |1\rangle_B
-e^{-iE\Delta t}|1\rangle_A |0\rangle_B,
\end{eqnarray}
where $\Delta t= t_A- t_B$.
In the proposed scheme we initially prepare quantum clock
pairs in the state $|\psi^-\rangle$ at a single site.
We let each quantum clock (labeled by $A$ or $B$)
departs and follows its own space-time trajectory and 
gather them again. Then we perform some (collective)
measurement on the quantum clocks and get information about
the proper-time difference
$\Delta t$.
(We do not consider the case where the terms differ by
$2n\pi$, $n$ is integer.)
As we see, the proper-time difference $\Delta t$ contributes
to the relative phase of the quantum clock pair.
Thus we can determine $\Delta t$ by measuring the relative
phase.
In other words,
{\it the difference $\Delta t$ of the proper-time
that each quantum clock experiences since they departed,
is accumulatedly
recorded in the relative phase of the non-degenerate
quantum clock pair in Eq. (\ref{d}),
which can be read out by collectively measuring the quantum clocks
at a single site.}
In the quantum clock synchronization \cite{jozs}, it is
required that the state remains in the initial one when
each clock has arrived at its own location. Namely
a condition 
that $t_A=t_B$ should be satisfied.
This condition
can be fulfilled by slow transportation of quantum clocks.
In this case, the relativistic effect is something
to be suppressed by a careful manipulation (slow
transportation) of quantum clocks.
In contrast, the proposed scheme
utilizes the (relative) phase rotation
of the state in Eq. (\ref{d}) when we measure the
relativistic time-dilation effect.

In the following, we explain why the proposed scheme would be
more precise in measuring the proper-time difference than the
scheme of two-separate-quantum-clocks. In the latter, proper-times
of two separate quantum clock which have travelled through
different space-time trajectories are compared to estimate the
difference between them.

Roughly speaking, in the proposed scheme
the (relative) phase corresponding to the proper-time
difference become stationary while measurement is done.
Thus the proper-time difference
can be more accurately measured in the proposed scheme.

Let us consider  simple measurement models and then,
using these, discuss on the advantage of the entangled scheme.

In separate quantum clocks scheme,
quantum clocks are initially prepared in the state
$|\bar{0}\rangle= |0\rangle+|1\rangle$.
(In this notation,
$|\bar{1}\rangle=|0\rangle-|1\rangle$.)
In order to measure the phase,
we perform, for example, a measurement $\hat{S}_x$
composed of two projection operators
$|\bar{0}\rangle \langle \bar{0}|$
and $|\bar{1}\rangle \langle\bar{1}|$.
The measurement  $\hat{S}_x$ on $\alpha$-th quantum clock can be
done
by applying the following
interaction Hamiltonians ${\bf H}_\alpha^I$ between
each quantum clock
and an ancillary quantum bit for a time width $\delta t$
\cite{neum,pre2,bohm}.
\begin{eqnarray}
\label{e}
{\bf H}_A^I=
{\bf \sigma}_x \otimes I \otimes  F,
\hspace{5mm}
{\bf H}_B^I =
I \otimes \sigma_x \otimes  F,
\end{eqnarray}
where $I$ is the identity operator and
$F$ is a certain operator that acts on ancillary quantum bit.
Here the time width $\delta t$ is inevitably finite because it 
describes real physical processes.
Let us consider $A$-quantum clock. (The same thing can be said 
for $B$-quantum clock.)
The total Hamiltonian ${\bf H}_A^T$ can be written as
\begin{eqnarray}
\label{f}
{\bf H}_A^T &=& {\bf H}_A + g(t) \cdot {\bf H}_A^I
\nonumber\\
            &=& E \sigma_z \otimes I \otimes  I
                 + g(t) \cdot {\bf \sigma}_x \otimes I \otimes  F,
\end{eqnarray}
where $g(t)$ is a Gaussian-like function
that is peaked at the time when measurement is performed and
whose  half-width is $\delta t$.
During the measurement, the prepared quantum clock rapidly 
rotates between
$|\bar{0}\rangle$ and $|\bar{1}\rangle$
due to its own Hamiltonian ${\bf H}_A$.
Since  $[{\bf H}_A$, ${\bf H}_A^I]  \neq 0$ 
($[C,D]=CD-DC$) and $\delta t \neq 0$,
the measurement result is inevitably  affected
by the evolution due to ${\bf H}_A$.
(To suppress this effect, it is assumed that either
${\bf H}_A=0$ or  $\delta t \rightarrow 0$ in many cases
\cite{neum,pre2,bohm}.)
Namely, during the measurement interval $\delta t$, the phase to
be measured is rotated $ 2 \pi \delta t/T  (T = \pi/E)$.
Thus, if the measurement time-width $\delta t$ is
non-negligible comparing with the
period of rotation $T$,
the result of the measurement would
be an average of phases of all states
in which the prepared quantum clock stays during a full rotation.
In this case, therefore, the measurement would fail or at least be
largely uncertain, if $\delta t > T$.
Now let us consider the proposed scheme. Here the pair of quantum clocks
are prepared in the state $|\Psi^- \rangle$ in Eq. (\ref{a}).
In order to measure the relative phase in Eq. ({\ref{d}) later,
we perform, for example, a measurement with
the following interaction Hamiltonian.
\begin{eqnarray}
\label{g}
{\bf H}^I = \vec{\sigma}_T^2 \otimes F,
\end{eqnarray}
where $\vec{\sigma}_T= \vec{\sigma} \otimes I+
I \otimes \vec{\sigma}$ and $\vec{\sigma}= (\sigma_x, \sigma_y,
\sigma_z)$.
This corresponds to total-spin measurement  \cite{saku}
in the case where the quantum clocks are
spin-1/2 states.
Similarly to above separate-case,
the total Hamiltonian ${\bf H}^T$ is given by
\begin{eqnarray}
\label{h}
{\bf H}^T &=& {\bf H}_A+  {\bf H}_B+ g(t) \cdot {\bf H}^I.
\end{eqnarray}
However, since
$[{\bf H}^I, {\bf H}_A+ {\bf H}_B]=0$ here, we can safely measure
the quantity corresponding to $\vec{\sigma}_T^2$ \cite{neum,pre2}.
Then let us decompose the state in Eq. (\ref{d}) as
\begin{eqnarray}
\label{i}
&& e^{iE\Delta t} |0\rangle_A |1\rangle_B
     -e^{-iE\Delta t}|1\rangle_A |0\rangle_B.
\nonumber\\
&=&  \cos (E\Delta t)  |\Psi^-\rangle+ i \sin (E\Delta t)   |\Psi^+\rangle,
\end{eqnarray}
where  $|\Psi^-\rangle$ and  $|\Psi^+\rangle$
are eigenstates of $\vec{\sigma}_T^2 $ with eigenvalues
$0$ and $1$, respectively.
($|\Psi^+ \rangle=
|0\rangle_A |1\rangle_B +|1\rangle_A |0\rangle_B$.)
However, the relative phase $E\Delta t$ does not evolve at the
measurement stage and thus finiteness of $\delta t$ does not matter.
Therefore,
by measuring $\vec{\sigma}_T^2$ with ${\bf H}^I$ in
Eq. (\ref{g}), for example, we can obtain coefficients in
Eq. (\ref{i}) and then calculate $\Delta t$.

Now let us continue to discuss on advantage of the proposed scheme.
It is clear that accuracy of the two-separate-clocks scheme
is limited by
the uncertainty
of each separate clock.
That is, the proper-time difference cannot be measured
more accurately than the uncertainty of each clock's time.
(Here we assume the period $T= \pi/E$ of quantum clock's phase
rotation is a constant, which is
equivalent to assuming complete shielding of quantum clocks from
environments.
Incompleteness of the shielding might be the limiting factor for
quantum clocks in some cases. In this case, phase-uncertainty
improvement by the proposed scheme would not be of
much importance. Thus
what we consider is the case where such complete-shielding problem
is overcome by certain methods.
Similar thing can be said to the efficient quantum clock
transport scheme \cite{chua} which improves phase-uncertainty.
However, even in this case, the phase
uncertainty would limit the accuracy of quantum clocks.)

The accuracy of a quantum clock is roughly
proportional to the product of the period $T$ of phase rotation and
the uncertainty in the  phase measurement $\delta \phi$.
The uncertainty of the phase may be due to the inherent
statistical behavior of quantum states
(i.e., the results of the phase-measurement
form a statistical
distribution given by quantum mechanical formula)
and inherent finite time width such as $\delta t$ of 
the function $g(t)$
in Eqs. (6) and (8)
involved with phase-measurement.
If we employ many quantum clocks, we can reduce the phase
uncertainty; roughly
$2^{2n}$ number of quantum clocks allow us to estimate
$n$ bits of the phase \cite{chua}.
When the number of quantum clocks is given, one may futher improve
the accuracy by decreasing $T$ i.e. by increasing
the speed of phase rotation.
(We can consider imporvement of accuracy by other method, namely
by optimizing the initial states \cite{buze}.) 
However, this method has its own limitation as the following.
The faster a phase rotates
the larger the phase uncertainty  would become, since
the phase makes wider angle of rotation during the measurement:
in real experiment our measurement-results for the phase
would inevitably correspond to
the phases during the inherent finite time width such as
$\delta t$  of the function $g(t)$ in Eqs. (6) and (8),
not that of an instance.
Thus the measurement-results for phases
make a broader statistical
distribution than in the case where
$\delta t$ is zero, thus increasing
the phase uncertainty for a given number of quantum clocks.
In particular, the $\delta t$'s broadening effect would
be considerable when $T$ become comparable with $\delta t$.
Moreover when $T$ become smaller than
$\delta t$, due to cyclic property of phase,
the phase uncertainty would be rapidly maximized
so that it would become impractical
to determine the phase.
(Therefore an optimal accuracy of quantum clocks would be
obtained by employing quantum bit systems with
a certain $T$ of intermediate value.)

In contrast,
in the proposed scheme the relative phase
in Eq. (\ref{d}) to be measured does not rotate
while measurements for the relative phase, for example,
the $\sigma_T^2$ measurement, are being performed.
Thus the inherent finite time width $\delta t$ involved
with phase-measurement does
not matter in the proposed scheme.
On the other hand, we can see
that accuracy of the proposed scheme
is given by a product of each separate
quantum clock's period $T$
of phase rotation
and uncertainty in the
relative phase $\delta \phi$. (Here note that $T$ is not
the period of the relative phase's rotation but
fast rotation of each quantum clock's phase.)
Thus we can improve the accuracy, by
decreasing $T$ as we like without increasing
$\delta \phi$ in the proposed scheme.

Now let us consider
quantum clocks whose phase rotation can be
turned on or off, as we like by some operation.
For example, spin precession of particles by applied
magnetic field can used as quantum clocks. Here we can
make the clocks
turned on (off) by applying nonzero (zero) magnetic field.
In this case, accuracy of the two-separate-clocks scheme
also would not be limited by the $\delta t$,
since we may turn off both
clocks when we are measuring them.
However, in this case
the magnetic field instead must be precisely
controlled to the level of required accuracy of the scheme,
which would be a much more difficult task than attaining
the required accuracy with two naturally given energy
eigenstates as in ordinary quantum clocks scheme.

The assumption that environmental effects can be
efficiently
removed is crucial for the success of the proposed scheme.
One may ask that if such efficient shielding is possible
or accurate quantum clocks can be obtained
then why we need the
entangled quantum clocks scheme. However, as noted above,
efficient shielding would not directly
guarantee accurate quantum
clocks, due to uncertainty in phase measurement.
Overcoming the phase uncertainty would become particularly
important in precise measurement of
the difference of proper-times.
The proposed scheme is advantageous
in that it is not limited by the inherent time width
$\delta t$ involved with phase-measurement,
in overcoming the phase uncertainty.

Let us now consider the general case where $E_\alpha$'s are
time-dependent and thus are not the same.
$E_\alpha$ may be time-dependent due to
either interaction
with environments or acceleration that each quantum clock suffers
during the round trip in space-time.
As previously,
here we assume that the environmental effects
can be removed by shielding the quantum clocks from
environment.
It is not well known yet how quantum clocks are
perturbed by acceleration except for the fact that
the effect would be very small \cite{misn}.
The condition that $E_A=E_B=E$ we assumed previously
also implies that acceleration effects can be removed
by some methods, e.g., careful choice of
the system to be used as quantum clocks or simply
making the acceleration very small.
Now we consider the case where
the acceleration effects
on the quantum clock's time evolution
is non-negligible.
The phase difference
$\Delta \phi= \int_0^{t_A}
E_A (t^{\prime}_A) dt^{\prime}_A
-\int_0^{t_B} E_B (t^{\prime}_B) dt^{\prime}_B$
that we would
measure in the proposed scheme is the combined results
of the relativistic and the acceleration effects, which cannot
be discriminated from each other in the measurement
result. Nevertheless, we can
still utilize the proposed scheme for measuring the proper-time
difference by making each quantum clock to experience the same
acceleration effect while following different path in
space-time. For example, we can consider the following case
almost similar to the twin-paradox experiment \cite{tayl}.
One (the other) party makes a short (long) trip.
However their
accelerating sections in the space-time trajectory
are the same with each other.
In this case the acceleration effect cancels
with each other and thus we can estimate pure
relativistic time-dilation effect from the measurement
result. Let us consider another
example, gravitational time-dilation \cite{poun,misn} where
we can also make the acceleration effects to cancel
with each other.
First prepare the two parties of entangled quantum clocks
at one site in constant gravitational field.
Then lift
both of them to a higher place and then bring down one
party to the original place. After waiting for a long time,
bring down the other party with the same magnitude of
acceleration and velocity as the first one.
Then by measuring the phase difference we can estimate
the gravitational time dilation.
On the other hand, we can make use of the proposed scheme
for measuring the acceleration effect. Let one party to
be at rest and another party to make a trip with
some acceleration, as we do in the original
twin-paradox experiment \cite{tayl}.
Measure the phase difference
$\Delta \phi$ and calculate each party's proper-time using
formula of special relativity. The difference between them
is the acceleration effect.

The precision of the entangled quantum clocks scheme
is estimated to be order of period $T$, assuming
$\delta \phi \sim 1$. The period $T$ of hyperfine transition
is of order of $10 ^{-10}$ second. However, in the proposed
scheme a system with more rapidly rotating phase can be
employed since the phase become stationary while it is being
measured, as noted before. By choosing some quantum clocks whose
energy difference $E_\alpha$ is of an order of one electron
volt, one can obtain $T \sim 10^{-14}$ second. However, the
greater energy difference $E_\alpha$ becomes, the more
probable the higher energy state makes a  spontaneous
transition to the lower one in general. This problem may be
avoided by adopting some metastable states as quantum clocks,
although this problem would limit the accuracy of the
proposed scheme.

It is interesting to note that at least in principle
the proposed scheme may also be used to detect
time-dilation effect that
gravitational wave cause, in a setting similar to
a non-mechanical gravitational wave detector proposed by
Braginsky and Menskii \cite{brag,misn}; fix two component
quantum clocks on edges of a disk, making an angle $\pi/2$
with the origin of the disk. The disk is free-falling and is
constantly rotating in phase with a frequency component of
gravitational wave and the axis of rotation is pointing to
the source of the wave. Then one clock's time is constantly
dilated when compared with the other one's, due to gravitational field of
the wave. However, it still seems to be a formidable task at present
to detect gravitational wave using the proposed scheme.

In conclusion, we reported that the entangled pair of
non-degenerate quantum clocks
can be used as a specialized detector for precisely
measuring the difference of proper-times that each constituent
quantum clock experiences.
We described why the proposed scheme would be
more precise in the proper-time difference measurement
than a scheme in which readings of
two separate quantum clocks are compared.
Acceleration's effects on quantum clock's time evolution may be
non-negligible. In this case, we considered
some experiments where
the acceleration effect cancels with each other.
The proposed scheme can be used in precision test
of relativistic time dilation effects-the twin paradox
effect \cite{tayl}, gravitation time-dilation
\cite{poun,misn}, and
possibly time-dilation due to the gravitational wave.

\acknowledgments
This work was supported by the Korean Ministry of Science
and Technology through the Creative Research Initiatives
Program under Contract No. 99-C-CT-01-C-35.
We are very grateful to Dr. Jinhyoung Lee, Dr. Andi Song,
and Dr. Xiang-Bin Wang
for valuable comments.

\end{document}